\documentstyle[aps,12pt]{revtex}

\draft

\input psfig.tex

\newcommand{\be}{\begin{eqnarray}}
\newcommand{\ee}{\end{eqnarray}}

\begin{document}

\title{Fractional Calculus as a Macroscopic Manifestation of Randomness}
\author{$^{1, 2, 3}$P. Grigolini, $^{1}$A. Rocco and $^{1}$B.J. West}
\address{$^{1}$Center for Nonlinear Science, University of North
Texas, P.O. Box 5368,
Denton, Texas 76203-5368}
\address{$^{2}$Dipartimento di Fisica dell'Universit\`{a} di Pisa, Piazza
Torricelli 2, 56100, Pisa, Italy}
\address{$^{3}$Istituto di Biofisica del Consiglio Nazionale delle
Ricerche,
Via S. Lorenzo 26, 56127 Pisa, Italy}
\date{\today}

\maketitle

\begin{abstract}
We generalize the method of Van Hove so as to deal with the case of
non-ordinary statistical mechanics, that being phenomena with no time-scale
separation. We show that in the case of ordinary statistical mechanics, even
if the adoption of the Van Hove method imposes randomness upon Hamiltonian
dynamics, the resulting statistical process is described  using normal
calculus techniques. On the other hand, in the case where there is no
time-scale separation, this generalized version of Van Hove's method not
only imposes randomness upon the microscopic dynamics, but it also transmits
randomness to the macroscopic level. As a result, the correct description of
macroscopic dynamics has to be expressed in terms of the fractional calculus.
\end{abstract}



\section{Introduction}

The physical paradigm of statistical physics is Brownian motion, which
involves diffusion, dissipation and the fluctuation-dissipation relation
tying the two together. The dynamical model of this process was provided by
Langevin in 1908 using a stochastic differential equation. In spite of this
long history it seems apparent from the nature of randomness that such
macroscopic stochastic equations are incompatible with the continuous and
differentiable character of microscopic Hamiltonian dynamics. However, it is
widely believed that Brownian motion can be rigorously derived from the
totally deterministic Hamiltonian models of classical mechanics. Part of the
reason for this conviction has to do with the wide use made in literature of
Van Hove's method \cite{vh,zwa,nota}. In one form or another, many of the
attempts currently made to establish a unified view of mechanics and
thermodynamics \cite{pri} can be traced back to the method of Van Hove. The
result of this method depends on whether we adopt the Heisenberg
perspective, corresponding to the time evolution of observables, or the Schr\"{o}dinger perspective, corresponding to the time evolution of the
Liouville density. In the former case, the usual outcome is the derivation
from mechanics of an ordinary Langevin equation. In the latter case, the
adoption of the Van Hove method yields master equations. We focus here
especially on the conventional diffusion equation, with the diffusion
process described by a second-order spatial derivative.

In the Heisenberg perspective, after averaging over an ensemble of
realizations of the stochastic force, the relaxation process is described by
an exponential function. In the Schr\"{o}dinger perspective, the
mathematical representation of the diffusion process is given, as we have
said, by a second-order spatial derivative of a density function. Therefore,
the mathematical description rests on either ordinary analytical functions
(exponential functions) describing the dynamics, or on conventional
differential operators (second-order derivatives) describing the phase space
evolution. This is probably the reason why there is no mention of the fact,
and indeed no perception, that the Van Hove method is equivalent to
introducing stochastic dynamics, namely discrete processes and unpredictable
non-differentiable jumps, in the dominion of continuous and differentiable
Hamiltonian dynamics.

Actually the differentiable nature of the macroscopic picture is, in
a sense, a natural consequence of the microscopic randomness \cite{GASPARD98,SPOHN98}, and of the related non-differentiability as well, due to the key
role of the central limit theorem. Recall
that in the central limit theorem the quantities being added together are
statistically independent, or at most weakly dependent, in order for the
theorem to be applicable and Gaussian statistics to emerge. Once a condition
of time-scale separation is established, in the long-time limit the memory
of the non-differentiable character of microscopic dynamics is lost, and
Gaussian statistics result. This also means that use can be made again of
ordinary differential calculations on the macroscopic scale, even if the
microscopic dynamics are incompatible with the adoption of ordinary calculus
methods.

On the other hand, in the case where a time-scale separation between
the macroscopic and the microscopic level of description does not exist, the
non-differentiable nature of the microscopic dynamics is transmitted to
the macroscopic level. An illuminating example is given by the paper of Ref.
\cite{alle}, which shows that a diffusion process generated by a fluctuation
with no time-scale at the macroscopic level generates a diffusion process
well described by a fractional Laplacian \cite{sew,wg}. The paper of Ref. 
\cite{alle}
addresses also the intriguing problem of making compatible a L\'{e}vy 
process, which has an infinite
second moment, with the dynamical approach to diffusion. The dynamical
approach to diffusion rests on steps of finite length, and consequently
results in  finite second moments. This problem has been addressed in a
variety of ways \cite{Fogedby94,MS94}, ranging from taking into account the
finite size $L$ of the sample, within which the L\'{e}vy flight takes place
\cite{Fogedby94}, to the case where the probability density is truncated 
\cite{MS94}.
We rather follow the prescriptions of \cite{alle}, which is a form of L\'{e}vy
walk, where the individual jumps are not instantaneous and involve a time
cost, thereby making it possible for us to adopt a diffusion picture with
finite moments. All this has been discussed in Ref. \cite{alle}, and here the
results of Ref. \cite{alle} are made compatible with the infinite moments of
the L\'{e}vy statistics by means of the generalized Van Hove method.  Another
case where the adoption of a fractional calculus is made necessary concerns
the generalization of the exponential form of relaxation according to the
recent prescriptions of Nonnenmacher and co-workers \cite{nonne,glockle}, involving the
concept of a fractional time derivative. The main aim of this paper is to
prove that the fractional derivative, both the space fractional derivative
of Ref. \cite{alle} and the time fractional derivatives of Refs. \cite{nonne,glockle} are made
compatible with Hamiltonian deterministic dynamics by means of a
generalized version of Van Hove's method.

In fact, if we make the traditional classical assumption that the microscopic
dynamics follow the Newton prescription, we also have to address the problem
of how to make this prescription compatible with randomness. The solution to
this problem requires that an extension of the Van Hove method be
discovered. This naturally yields the working hypothesis that the fractional
derivatives, currently used to describe macroscopic transport processes \cite{sew,nonne,glockle} can be regarded as the macroscopic manifestation, in the
absence of time-scale separation, of either non-differentiable microscopic
dynamics, an assumption that would violate the applicability of Hamiltonian
dynamics to this domain, or of a Hamiltonian description that loses
differentiability through a kind of filtering described by a generalized Van
Hove approximation. In other words, just as the Van Hove method makes
Brownian motion compatible with Hamiltonian dynamics, a generalized Van Hove
method is used here to make the macroscopic fractional calculus compatible
with microscopic Hamiltonian dynamics. The main difference with the case of
ordinary statistical mechanics is that in this non-ordinary case the
non-differentiable nature of microscopic dynamics, either natural or forced
by the adoption of the Van Hove prescription, is transmitted to the
macroscopic level where it takes the shape of fractional derivatives. The
main purpose of this paper is to substantiate with convincing arguments this
working assumption.

The outline of the paper is as follows. In Section II we express the Van
Hove method in a form, equivalent to the ordinary method, but more
convenient for the generalization that we plan to develop in this paper. In
Section III we show that in the Heisenberg picture the generalization of the
Van Hove method results in the fractional derivatives currently used by
Nonnenmacher and co-workers to study polymer dynamics. Section IV and
Appendix 1 are devoted to the Schr\"{o}dinger picture, and show that the Van
Hove method in this picture leads to the fractional derivative introduced by
Seshadri and West \cite{sew} (see also West and Grigolini \cite{wg}) to
describe L\'{e}vy processes. Section V is devoted to concluding remarks.

\section{The Van Hove limit: an example from a specific model}

As pointed out in the Introduction, the Van Hove limit \cite{vh} of a
microscopic process turns out to be a key ingredient for the derivation of
statistical mechanics from microscopic dynamics. We review this limit and
how it works in the case of a specific model rather than in general.

Let us start by recalling the meaning of this limit. Consider the 
integro-differential equation 
\be
\frac{df(t)}{dt}=-\lambda ^{2}\int_{0}^{t}K(t-t^{\prime })f(t^{\prime})dt^{\prime }, \label{effe}
\ee
where $f$ is some quantity of interest, $K$ is a memory kernel and $\lambda $
is a parameter. Equation (\ref{effe}) is a typical non-Markovian equation
obtained in studying physical systems coupled to an environment, and whose
environmental degrees of freedom have been averaged over. In this case the
parameter $\lambda $ can be regarded as the strength of the perturbation
induced by the environment on the system of interest.

In the literature, wide use is made of the Markov approximation
\cite{pgri}, which replaces the integro-differential equation (\ref{effe}) with the rate equation
\be
\frac{df(t)}{dt}=-\left( \lambda ^{2}\int_{0}^{\infty }K(t^{\prime
})dt^{\prime }\right) f(t). \label{effem}
\ee
The Van Hove limit \cite{zwa} consists in making the limit $\lambda \rightarrow 0$, $t \rightarrow \infty$ in such a way that the product $\lambda^2 t$ is kept constant. That is, setting $x = \lambda^2 t$ and $F(x) = f(t)$, Eq.(\ref{effe}) becomes:
\be
\frac{dF(x)}{dx}=-\int_{0}^{x/\lambda^2}K(t^{\prime})F(x - \lambda^2 t^{\prime}) dt^{\prime }. \label{reee}
\ee
Now, the adoption of the limit
\be
{\rm const} \;\;\; = \;\;\; x \;\;\; = \lim_{
\begin{array}{c}    
\lambda \rightarrow 0 \\
t \rightarrow \infty 
\end{array}}
\lambda^2 t, \label{vanho}
\ee
makes it possible for us to replace the time convoluted form in Eq.(\ref{effe}) with
\be
\frac{df(t)}{dt} = -\lambda^2 \tau f(t), \label{fdec}
\ee
where
\be
\tau = \int_0^{\infty} dt K(t).
\ee
Eq.(\ref{fdec}) gives an exponential solution, the same as that obtained by means of the Markov approximation (\ref{effem}).

Taking the limit $\lambda \rightarrow 0$ corresponds to assuming that the
coupling of the system to the environment is weak, while the limit $ t\rightarrow \infty $ means that the observation time is much larger than
other temporal scales present in the system. Specifically this time must be
larger than the microscopic time $\tau $. This remark allows us to
reformulate the Van Hove limit in a slightly different way, more suitable
for our purposes. First, instead of taking the $t\rightarrow \infty $ limit,
we shall take the limit $\tau \rightarrow 0$. Also, we shall replace the
limit $\lambda \rightarrow 0$ with the equivalent limit $V\rightarrow \infty 
$, where $V$ is a coupling constant, to be specified in the following. The
quantity to be kept constant in carrying out the limit is just the product $V^{2}\tau $. Notice that connecting $V$, $\tau $ and $\lambda $ as $\lambda
=V\tau $ makes it possible to keep $V\ll \tau ^{-1}$ (in such a way that $\lambda \ll 1$) and at the same time to set $V^{2}\rightarrow
\infty $ so as to ensure $V^{2}\tau \rightarrow {\rm const}$.

The meaning of Eqs. (\ref{reee}) to (\ref{fdec}) can be illustrated by 
adopting this
equivalent perspective and can be shown to be a way of disregarding the
time evolution of the system at both very short and very long times, where
the deviations from the exponential relaxation show up \cite{fonda}. In the
Hamiltonian case
the relaxation cannot be rigorously exponential \cite{lee} thereby preventing a
satisfactory connection between microscopic dynamics and stochastic physics
\cite{Fox77}. We now show this benefic effect of the Van Hove method with the
help of an illustrative example, where the origin of Brownian motion stems
from the average over an initial statistical distibution \cite{SPOHN98} rather
than from chaos \cite{GASPARD98}. Consider a chain of
$2N+1$ linear harmonic oscillators all with equal spring constants and
described by the  Hamiltonian 
\be
H=\sum_{i=-N}^{N}\frac{p_{i}^{2}}{2m_{i}}+\frac{k}{2}\sum_{i=-N}^{N}(q_{i+1}-q_{i})^{2}, \label{hami}
\ee
where $m_{i}=m$ if $i\neq 0$ and $m_{0}=M$. Vitali and Grigolini \cite{vitali} prove
that the correlation function for the momentum $p_{0}$ of the system of
interest 
\be
\Phi _{0}(t)\equiv \frac{\langle p_{0}p_{0}(t)\rangle }{\langle
p_{0}^{2}\rangle }
\ee
satisfies the following integral equation 
\be
\dot{\Phi}_{0}=-\Delta _{1}^{2}\int_{0}^{t}\Phi _{1}(t-t^{\prime })\Phi_{0}(t^{\prime })dt^{\prime }, \label{pero}
\ee
where 
\be
\Delta _{1}^{2}=2k/M
\ee
and $\Phi _{1}(t)$ represents the correlation function of the stochastic
force in the corresponding generalized Langevin equation (see Section III).
The parameter $\Delta _{1}$ plays the same role as that of the earlier
mentioned coupling constant $V$, while $\tau $ is given in this specific
model by the expression: 
\be
\tau \equiv \int_{0}^{\infty }\Phi _{1}(t)dt=\sqrt{\frac{m}{k}}.
\ee
In the long-time region, the solution to Eq.(\ref{pero}) is given by \cite
{vitali} 
\be
\Phi _{0}(t) &=&\frac{1-\mu }{1-2\mu }\exp \left( -\frac{\mu \omega _{0}t}{\sqrt{1-2\mu }}\right) + \frac{2\mu }{\pi }\int_{0}^{\infty }\frac{\sin (x\omega
_{0}t)\sqrt{x^{2}-1}}{(1-2\mu )x^{2}+\mu ^{2}}dx \nonumber \\
&\approx& \frac{1-\mu }{1-2\mu }\exp \left( -\frac{\mu \omega _{0}t}{\sqrt{1-2\mu }}\right) +\frac{\mu }{(1-\mu )^{2}}\sqrt{\left( \frac{2}{\pi (\omega _{0}t)^{3}} \right) }\sin \left( \omega _{0}t-\frac{\pi }{4}\right) ,  \label{pero2}
\ee
where 
\be
\mu =m/M
\ee
and 
\be
\omega _{0}^{2}=4k/m.
\ee
The asymptotic approximation (\ref{pero2}) shows that the momentum 
autocorrelation function consists of the
sum of an exponentially decaying term and a non-exponential oscillatory
contribution that decays as an inverse power law in time. Notice that $\omega _{0}$ is a Debye-like cutoff frequency that is related to the
microscopic time-scale $\tau $ as $\tau =2/\omega _{0}$. The general form of
this solution had also been observed and discussed by Zwanzig \cite{zwa}

Now, the Van Hove limit expressed in terms of the parameters of the model
becomes: 
\be
\gamma \;\;\;\equiv \lim_{
\begin{array}{c}
\tau \rightarrow 0 \\ 
\Delta _{1}^{2}\rightarrow \infty 
\end{array}
}\Delta _{1}^{2}\tau \;\;\;=\lim_{
\begin{array}{c}
\mu \rightarrow 0 \\ 
\omega _{0}\rightarrow \infty 
\end{array}
}\omega _{0}\mu. \label{rel}
\ee
Notice that this limit can be realized assuming $m\rightarrow 0$ and $M\rightarrow 0$ as $m^{1/2}$. The closer to zero the mass $m$, the better
the physical condition $M\gg m$ is fulfilled (the better the macroscopic
description of the system of interest). Consequently, the oscillatory tails
in Eq.(\ref{pero2}) cancel and an exact exponential relaxation is recovered: 
\be
\Phi _{0}(t)=\exp (-\gamma t). \label{esp2}
\ee
The rationale for this result stays in the formal similarity of Eq.(\ref
{pero}) with Eq.(\ref{effe}). Letting, as mentioned before, $\lambda =\Delta
_{1}\tau $, it is a simple matter to verify that $\lambda \rightarrow 0$ as $m^{1/4}$ and the ordinary Van Hove limit (\ref{vanho}) can be applied. More
in general, we can assume that the relation (\ref{rel}) holds true
independently of the model, and we can regard the parameters $\mu $ and $\omega _{0}$ as free parameters of the theory. If we adopt this view, the exponential decay is recovered by making the Van Hove limit as described by the r.h.s. of Eq.(\ref{rel}).

In the following, we shall refer to the Van Hove limit, rather than using
the known version of Eq.(\ref{vanho}), as expressed in our non-conventional
form: 
\be
\gamma \;\;\;\equiv \lim_{
\begin{array}{c}
\tau \rightarrow 0 \\ 
\Delta _{1}^{2}\rightarrow \infty 
\end{array}
}\Delta _{1}^{2}\tau \equiv {\rm Lim_{VH}}\Delta _{1}^{2}\tau. \label{vhnc}
\ee

A final remark concerns the fact that the same result, Eq.(\ref{esp2}), could have
been obtained by applying the Markov approximation to Eq.(\ref{pero}). In that
case, the basic assumption would have been an infinite time-scale separation
between the microscopic time-scale $\tau $ and the macroscopic scale defined
as the inverse of the frequency $\Omega ^{2}=2k/M$, that is: 
\be
T=\sqrt{\frac{M}{2k}}.
\ee
Notice that the time-scale separation between the system of interest and
bath is rendered infinitely large exactly by the same limits as those
used in order to carry out the Van Hove limit: $m\rightarrow 0$ and $M\rightarrow 0$ as $m^{1/2}$. This demonstrates that the Van Hove limit and
the Markov approximation are essentially equivalent to one another.

\section{The Heisenberg picture: ordinary and fractional relaxation}

It is well known \cite{mori} that the generalized Langevin equation 
\be
\dot{v}=-\Delta _{1}^{2}\int_{0}^{t}\Phi _{1}(t-t^{\prime })v(t^{\prime})dt^{\prime }+f(t) \label{mori}
\ee
corresponds to the following hierarchy of correlation functions: 
\be
\dot{\Phi}_{i}=-\Delta _{i+1}^{2}\int_{0}^{t}\Phi _{i+1}(t-t^{\prime })\Phi
_{i}(t^{\prime })dt^{\prime }. \label{hye}
\ee
Eq.(\ref{pero}) is the $i=0$ case of Eq.(\ref{hye}).

In order to derive both normal and anomalous relaxation properties, we are
interested in making a non-trivial choice of the correlation function $\Phi
_{1}(t)$. However we also want our choice to be compatible with a completely
dynamical approach. Therefore we need to identify the conditions necessary
to assure that both these constraints are satisfied. First, by means of a Laplace Transforms and the convolution form of Eq.(\ref{hye}), it is easy to prove that $\Phi _{1}$ can be represented in the form
of a continued fraction: 
\be
\tilde{\Phi}_{1}(z)=\frac{1}{z+\displaystyle \frac{\Delta _{2}^{2}}{z+
\displaystyle \frac{\Delta _{3}^{2}}{z+...}}},
\ee
where $\tilde{\Phi}_{1}(z)$ denotes the Laplace Transform of $\Phi _{1}(t)$.
This structure is valid for $\Phi _{1}(0)=1$ and this is a first requirement
to fulfill. Then, we recall that the expansion parameters $\Delta
_{i}^{2}$ can be expressed in terms of the moments 
\be
s_{n}\equiv \frac{\langle f_{1}|(-L)^{n}|f_{1}\rangle }{\langle f_{1}|f_{1}\rangle }=\Phi _{1}^{(n)}(0),
\ee
where $L$ is the Liouvillian operator driving the time evolution of the
Liouville density and $|f_{1}\rangle $ is the first state in the Mori chain 
\cite{mori}. An elegant way of expressing the parameters $\Delta_{i}^{2}$ in terms of the moments $s_n$ has been established by Grigolini {\em et al.} \cite{ggps83}. This implies that $\Phi _{1}(t)$ must be infinitely differentiable. Finally, the symmetry properties of the
Liouvillian imply that also the condition
\be
s_{2n-1}=0 \label{star}
\ee
has to be fulfilled.

Therefore, we decide to focus our attention on the choice: 
\be
\Phi _{1}(t)=\frac{T^{\beta }}{(T^{2}+t^{2})^{\beta /2}}. \label{fi1}
\ee
Note that the moments $s_n$ are nothing but the $n$-th order time
derivatives of $\Phi_1(t)$. Thus, it is straightforward to prove via
successive time differentiation of (\ref{fi1}) that the odd moments vanish. 
This
is an important mathematical property necessary to make the relaxation
process compatible with Hamiltonian dynamics. An important example of
relaxation incompatible with Hamiltonian dynamics is in fact the
exponential relaxation: A case exhaustively discusses by Lee in his
brilliant 1983 paper \cite{Lee83}. This paper proves therefore the importance of
fulfilling the constraints of Eq. (\ref{star}) for a generalized Langevin 
equation
to be compatible with Hamiltonian dynamics. This condition is here
fulfilled by adopting the choice of Eq. (\ref{fi1}).

\subsection{Ordinary statistical mechanics}

Ordinary statistical mechanics can be recovered from the generalized
Langevin equation (\ref{mori}), using Eq.(\ref{fi1}) along with the
integrability condition on the power-law index, 
\be
\beta >1.
\ee
The microscopic time-scale $\tau $ is given in this case in terms of the
parameter $T$, 
\be
\tau =\int_{0}^{\infty }dt\frac{T^{\beta }}{(T^{2}+t^{2})^{\beta /2}}=\frac{\sqrt{\pi }}{2}\frac{\Gamma \left( \frac{\beta -1}{2}\right) }{\Gamma \left( 
\frac{\beta }{2}\right) }T.
\ee
Therefore, the Van Hove limit in the form (\ref{vhnc}) is achieved in the
limit $T\rightarrow 0$, 
\be
\gamma =\frac{\sqrt{\pi }}{2}\frac{\Gamma \left( \frac{\beta -1}{2}\right) }{\Gamma \left( \frac{\beta }{2}\right) }{\rm Lim_{VH}}\Delta _{1}^{2}T.
\ee
This limiting procedure results in exponential relaxation for the
correlation function $\Phi _{0}(t)$ and allows us to safely interpret Eq.(\ref{mori}) as identical to the ordinary Langevin equation, 
\be
\dot{v}=-\gamma v+f(t).
\ee
This is the traditional result obtained using the Van Hove method.

\subsection{Non-ordinary statistical mechanics}

In the case where the power-law index of the correlation function is in the
interval
\be
0<\beta <1,
\ee
the non-integrability of the correlation function (\ref{fi1}) prevents us
from adopting the above approach. In the case $0<\beta <1$, we are forced
to look for a different procedure to go from a microscopic to a macroscpic
description of the system. This procedure can be derived in a natural way
from the original Van Hove limit. Let us consider the following limit, 
\be
Q\;\;\;=\;\;\;\lim_{
\begin{array}{c}
T\rightarrow 0 \\ 
\Delta _{1}^{2}\rightarrow \infty 
\end{array}
}\Delta _{1}^{2}T^{\beta }\equiv {\rm Lim_{GVH}}\Delta _{1}^{2}T^{\beta }.\label{gvh}
\ee
We shall refer to this limit as to the generalized Van Hove limit.

Adopting the ansatz (\ref{gvh}), and inserting Eq.(\ref{fi1}) into Eq.(\ref{mori}), we obtain 
\be
\dot{\Phi}_{0} &=&-{\rm Lim_{GVH}}\Delta _{1}^{2}T^{\beta }\int_{0}^{t}\frac{1}{(T^{2}+(t-t^{\prime })^{2})^{\beta /2}}\Phi _{0}(t^{\prime
})dt^{\prime }  \nonumber \\
&=&-Q\int_{0}^{t}\frac{1}{(t-t^{\prime })^{\beta }}\Phi _{0}(t^{\prime
})dt^{\prime }.  \label{vanh}
\ee
For dimensional reasons, it is convenient to write 
\be
Q\equiv V^{2}\tau ^{\beta }. \label{pio}
\ee
Notice that in general the correlation function is related to the waiting
time distribution $\psi (t)$ of the process under study as 
\be
\Phi _{0}(t)=1-\int_{0}^{t}\psi (t^{\prime })dt^{\prime }
\ee
and therefore Eq.(\ref{vanh}) can be rewritten: 
\be
\psi (t)=Q\int_{0}^{t}\frac{1}{(t-t^{\prime })^{\beta }}\Phi _{0}(t^{\prime})dt^{\prime }. \label{quadra}
\ee
Now, we want to compare this last expression with the fractional relaxation
equation obtained by Gl\"{o}kle and Nonnenmacher \cite{nonne}. The
fractional relaxation equation for a function $\Phi _{{\rm ML}}(t;\nu)$ is
given by
\be
\Phi _{{\rm ML}}(t;\nu)-\Phi _{{\rm ML}}(0;\nu)=\frac{1}{\tau ^{\nu }}\frac{%
d^{-\nu }}{dt^{-\nu }}\Phi _{{\rm ML}}(t;\nu) \label{fre}
\ee
where the symbol $\frac{d^{-\nu }}{dt^{-\nu }}$ denotes the fractional
integral (see Appendix 1), 
\be
\frac{d^{-\nu }}{dt^{-\nu }}f(t)=\frac{1}{\Gamma (\nu )}\int_{0}^{t}\frac{f(t^{\prime })dt^{\prime }}{(t-t^{\prime })^{1-\nu }}.
\ee
The solution of Eq.(\ref{fre}) is known \cite{glockle} and is given by the
so called Mittag-Leffler function,
\be
\Phi _{{\rm ML}}(t;\nu )=\Phi _{{\rm ML}}(0;\nu )\sum_{k=0}^{\infty }\frac{(-1)^{k}}{\Gamma (\nu k+1)}\left( \frac{t}{\tau }\right) ^{\nu k},
\ee \label{nabla}
which exhibits stretched exponential behavior at short times and inverse
power-law relaxation at long times.

Setting $\beta =1-\nu $, it becomes possible to compare the fractional
relaxation equation with Eq.(\ref{quadra}) to obtain 
\be
\psi (t)=-Q\Gamma (1-\beta )\tau ^{1-\beta }\left[ \Phi _{{\rm ML}}^{{}}(t;1-\beta )-\Phi _{{\rm ML}}^{{}}(0;1-\beta )\right]. 
\ee
where the superscript $1-\beta $ corresponds to the value for $\nu $ to
insert into the form of the solution (\ref{nabla}). Recalling also Eq.(\ref{pio}), we finally obtain for the waiting time distribution function 
\be
\psi (t)=-(V^{2}\tau )\Gamma (1-\beta )\left[ \Phi _{{\rm ML}}^{{}}(t;1-\beta )-\Phi _{{\rm ML}}^{{}}(0;1-\beta )\right] ,
\ee
which mantains the interesting properties of the Mittag-Leffler function 
earlier pointed out.

It is important to stress that the resulting analytical function
has been used by Gl\"{o}ckle and Nonnenmacher \cite{nonne91} to fit with a 
very remarkable accuracy the relaxation curves of stress experiments on glassy
material. This suggests that the dynamical randomness without time-scale
separation takes the shape of a time fractional derivative and becomes
experimentally detectable at the macroscopic level.

\section{Schr\"{o}dinger picture: Gaussian and L\'{e}vy diffusion}
Let us now consider the equation of motion for the one-dimensional
probability density $p(x,t)$: 
\be
\frac{\partial }{\partial t}p(x,t)=\langle \xi ^{2}\rangle_{eq}\int_{0}^{t}\Phi _{\xi }(t^{\prime })\frac{\partial ^{2}}{\partial x^{2}}p(x,t-t^{\prime })dt^{\prime }. \label{nongdiff}
\ee
This equation refers to the process: 
\be
\dot{x}=\xi.
\ee
and $\Phi _{\xi }(t)$ is the autocorrelation function of the $\xi -
$fluctuations. The integro-differential equation of motion for the
probability density is obtained from the Langevin equation by adopting the
Zwanzig projection operator method in the form discussed by Grigolini \cite
{gri} and it is exact under the following two conditions: (i) the dynamics
of $\xi $ is independent of that of $x$ and (ii) the system producing the $\xi -$fluctuations is a two-state system.

Notice that (ii) does not necessarily mean that the variable $\xi $ is
dichotomous. The case of anomalous diffusion generated by intermittent maps 
\cite{tref} is an illuminating example, where the change from a continuous
two-state fluctuating variable to the dichotomous case does not produce
significant effects on diffusion. From the point of view of our generalized
Van Hove limit, however, the difference in the two conditions is essential.
As we shall see in the next subsection, the replacement of $\xi $ with a
dichotomous variable with the values $W$ and $-W$ is an essential ingredient
of the generalized Van Hove method.

For reasons which will shortly become clear, it is convenient to write the
probability distribution at time $t-t^{\prime }$ in terms of the probability
distribution at time $t$ as 
\be
p(x,t-t^{\prime })=\int_{-\infty }^{\infty }F(x-x^{\prime },-t^{\prime})p(x^{\prime },t)dx^{\prime }, \label{tr1}
\ee
where the function $F$ is a propagator. Equation(\ref{nongdiff}), with the
non-trivial choice (\ref{fi1}), becomes  
\be
\frac{\partial }{\partial t}p(x,t) &=&\langle \xi ^{2}\rangle_{eq}\int_{-\infty }^{\infty }dx^{\prime }\int_{0}^{t}dt^{\prime }\frac{T^{\beta }}{(T^{2}+{t^{\prime }}^{2})^{\beta /2}} \frac{\partial ^{2}}{\partial x^{2}}F(x-x^{\prime
},-t^{\prime })p(x^{\prime },t).  \label{intprop}
\ee

We shall discuss both the case of ordinary diffusion, $\beta >1$, and the
case of anomalous diffusion, $0<\beta <1$.

\subsection{Ordinary statistical mechanics}

Let us assume that the power-law index in the autocorrelation function is
\be
\beta >1
\ee
and let us make the choice of a time-independent propagator 
\be
F(x-x^{\prime },-t^{\prime })=\delta (x-x^{\prime }). \label{tr2}
\ee
Equation(\ref{intprop}) then becomes 
\be
&&\frac{\partial }{\partial t}p(x,t) = \langle \xi ^{2}\rangle _{eq}\left\{ \int_{0}^{t}dt^{\prime }\frac{T^{\beta }}{(T^{2}+{t^{\prime }}^{2})^{\beta /2}}\right\} \frac{\partial ^{2}}{\partial x^{2}}p(x,t).  \label{intprop2}
\ee
The microscopic time-scale, $\tau ,$ of the autocorrelation function, $\Phi
_{\xi }(t),$ present in Eq.(\ref{intprop2}) is finite for $\beta >1$ and we can apply again the conventional Van Hove limit in the form 
\be
D\;\;\;=\lim_{
\begin{array}{c}
T\rightarrow 0 \\ 
\langle \xi ^{2}\rangle _{eq}\rightarrow \infty 
\end{array}
}\langle \xi ^{2}\rangle _{eq}T.
\ee
Substituting this expression into Eq.(\ref{intprop2}) we obtain the standard
diffusion equation 
\be
\frac{\partial }{\partial t}p(x,t)=D\frac{\partial ^{2}}{\partial x^{2}}p(x,t).
\label{mmm}
\ee

Note that the case where the correlation function $\Phi _{\xi }(t)$ is
exponential, although in conflict with the Hamiltonian constraints 
\cite{lee}, assists us in further interpreting the meaning of the Van Hove
limit. As shown in Appendix 2, using an exponential correlation function in Eq.(\ref{intprop}) along with Eq.(\ref{tr2}), reduces Eq.(\ref{intprop}) to the
telegraphic equation but ordinary diffusion can be still recovered by
making use of the Van Hove limit.

\subsection{Non-ordinary statistical mechanics}

Let us consider now the case where the power-law index is in the interval 
\be
0<\beta <1.
\ee
In this case, we define the generalized version of the Van Hove procedure as the recipe leading to the largest component of the diffusion process. Therefore, as a first step, we make the variable $\xi $ dichotomous,
by assigning to it the values $W$ and $-W$ with the basic condition that the
actual values of the true variable are included in the interval $[-W,W]$ (see Fig. 1). 
\begin{figure}[h] 
\centerline{{\psfig{figure=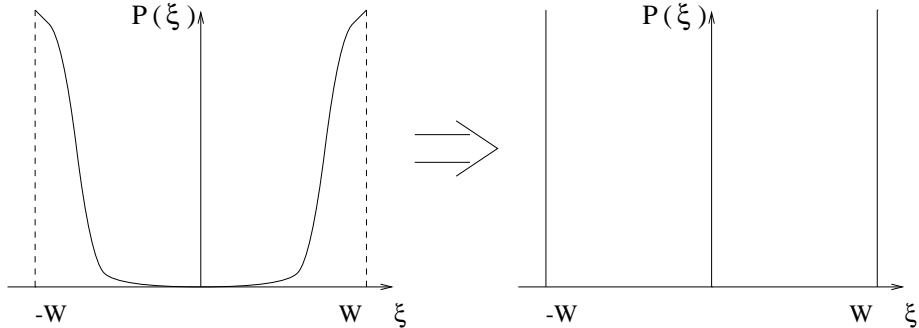,height=1.7in}}}
\caption{\small{Sketch of how the distribution $P(\xi)$ of the $\xi$ variable is modified so as to make the variable $\xi$ fully equivalent to a dichotomous variable.}}
\end{figure}
This leads us to assume for the form of the  propagator $F$, 
\be
F(x-x^{\prime },-t^{\prime })=\delta (Wt^{\prime }-|x-x^{\prime }|),
\ee
which is clearly time-dependent. Equation (\ref{intprop}) then becomes 
\be
&&\frac{\partial }{\partial t}p(x,t)=-\frac{W^{\beta +1}T^{\beta }}{2}\int_{-\infty }^{\infty }dx^{\prime }p(x^{\prime },t) \frac{\beta [W^{2}T^{2}+|x-x^{\prime }|^{2}]-\beta (\beta
+2)|x-x^{\prime }|^{2}}{[W^{2}T^{2}+|x-x^{\prime }|^{2}]^{\beta /2+2}}.
\label{intprop3}
\ee
It is evident that Eq.(\ref{intprop3}) differs from Eq.(\ref{intprop}) by a
correction term that disappears in the generalized Van Hove limit. The
generalized Van Hove limit is defined in this case as: 
\be
Q\;\;\;=\lim_{
\begin{array}{c}
T\rightarrow 0 \\ 
W\rightarrow \infty 
\end{array}
}T^{\beta }W^{1+\beta }
\ee
and noticing that this implies $W^{2}T^{2}\rightarrow 0$ as $W^{-2/\beta }$,
Eq.(\ref{intprop3}) can be rewritten 
\be
\frac{\partial }{\partial t}p(x,t)=\frac{\beta (\beta +1)}{2}Q\int_{-\infty}^{\infty }dx^{\prime }\frac{p(x^{\prime },t)}{|x-x^{\prime }|^{\beta +2}}. \label{sw}
\ee
This form of diffusion equation coincides with the West-Seshadri equation \cite{alle,sew} for a centro-symmetric L\'{e}vy process. The fact that Eq.(\ref{sw}) allows for a solution in the form of a L\'{e}vy process becomes
apparent by taking its Fourier transform. Letting 
\be
\mu =\beta +1\;\;\;\Rightarrow \;\;\;1<\mu <2 \label{condi}
\ee
and using \cite{grad}, we obtain for the Fourier transform of an inverse
power law 
\be
{\cal F}\left( \frac{1}{|x|^{1+\mu }};k\right) =\sqrt{\frac{2}{\pi }}\Gamma (-\mu )|k|^{\mu }\cos \left( -\frac{\pi \mu }{2}\right) .
\ee
Therefore, making use of the Convolution Theorem for Fourier Transforms, Eq.(\ref{sw}) becomes: 
\be
\frac{\partial }{\partial t}\phi (k,t)=-b|k|^{\mu }\phi (k,t). \label{swf}
\ee
Here, $\phi (k,t)$ is the Fourier Transform of $p(x,t),$ that is, the
characteristic function, and the parameter $b$ is given by 
\be
b=\frac{\beta (\beta +1)}{\sqrt{2\pi }}Q\Gamma (-\mu )\left| \cos \left(\frac{\pi \mu }{2}\right) \right| >0.
\ee
The solution to Eq.(\ref{swf}), with the initial condition $\phi (k,t=0)=1$,
necessary for the inverse Fourier transform to be defined as a probability
density, is written as 
\be
\phi (k,t)=e^{-b|k|^{\mu }t},
\ee
which indeed corresponds to the definition of the characteristic function
for a L\'{e}vy process.

Equation(\ref{sw}) can also be cast in the form of a fractional differential
equation \cite{wg}
\be
\frac{\partial }{\partial t}p(x,t)=(-i)^{\mu }b{\cal D}^{(\mu )}p(x,t), \label{frat}
\ee
where ${\cal D}^{(\mu )}$ is a proper definition of fractional derivative
(see Appendix 1). We see, in conclusion, that with the adoption of the generalized Van Hove method the standard diffusion equation yielded by the ordinary Van Hove method is expressed in terms of a fractional derivative. More in general, we
expect that the ordinary Fokker-Planck equation can be replaced by a
generalized expression resting on fractional derivatives \cite{wg}.

In the next Section we shall give more support to our conviction
that the generalized form of Van Hove method, here adopted to derive the
spatial version of fractional derivatives of (\ref{frat}) is intimately 
related to
the L\'{e}vy-Gnedenko theorem \cite{gk54}. We limit ourselves to point out that the
adoption of fractional calculus to deal with processes of anomalous
diffusion is becoming more and more popular, and we quote for the
interested reader Refs. \cite{MN98,MKS98}, whose results, however, must be
compared to the conclusion of Ref. \cite{ZK93} as well as to those of this
Section with some caution \cite{note}

\section{Concluding remarks}

The adoption of the Van Hove limit is essentially a sophisticated way of
making the Markov approximation. The Markov approximation, in turn,
establishes the physical condition necessary to make Hamiltonian dynamics
compatible with stochastic physics. However, when this method is applied to
microscopic dynamics to derive ordinary statistical mechanics, there is no
clear perception of establishing dynamic properties inconsistent with
Hamiltonian dynamics. This is so because even if the correlation functions
are made exponential by forcing the Markov approximation into microscopic
dynamics, so as so to become incompatible with both classical \cite{lee} and
quantum \cite{fonda} mechanics, they are still differentiable functions.

From the point of view of a single trajectory, the realization of the
Brownian condition implies, from a rigorous mathematical point of view, the
breakdown of the condition of differentiability. However, even in this case
the perception of a conflict with Hamiltonian dynamics is blurred by the
adoption of a statistical perspective. Within the Schr\"{o}dinger-like
picture, namely the picture where we observe the motion of an ensemble of
trajectories rather than that of a single trajectory, ordinary diffusion is
produced by a dynamical operator proportional to a second-order spatial
derivative. Again, a condition of ordinary differentiability is ensured! In
conclusion, the existence of a time-scale separation between microscopic
dynamics and the experimental observation, usually made at the macroscopic
level, ensures that the results of the observation process can be predicted
by means of theoretical prescriptions based on ordinary mathematical
procedures resting on the differentiability assumption.

A totally different condition is generated if the time-scale separation is
not adequate. We have separately discussed the Heisenberg and the Schr\"{o}dinger picture. From the results of Section III, devoted to the Heisenberg
picture, we see that when the condition of time-scale separation is insured,
the ordinary Van Hove method can be applied, and standard exponential
relaxation follows. In the absence of the condition of time-scale separation
the method of Van Hove must be generalized and this, in turn, yields a
generalization of the exponential relaxation, a condition that according to
Gl\"{o}ckle and Nonnenmacher \cite{nonne,glockle} turns out to be very efficient for describing such non-standard physical processes as polymer dynamics.

From within the Schr\"{o}dinger picture, discussed in Section IV, the lack
of a time-scale separation and the consequent requirement for a
generalization of the Van Hove method, yields the striking replacement of
the ordinary Laplacian operator with the fractional Laplacian discussed in
Appendix 1. We have to point out that even in this case randomness has a
very subtle origin, implying a departure from the differentiable condition
of Hamiltonian dynamics, and the replacement of the continuous variable
responsible for microscopic fluctuations with a dichotomous variable. This
process, changing a two-state physical dynamics, compatible with a
Hamiltonian picture, into a non-differentiable process is clearly described
by the sketch of Fig. 1. We firmly restate here that the replacement of a
continuous variable picture with a discrete representation is done also in
the case of ordinary statistical mechanics, where this way of forcing
randomness within differentiable dynamics does not imply a departure from
differentiable dynamics at a statistical level. When the condition of
time-scale separation is not available, the Van Hove procedure of forcing
microscopic dynamics to become random has a striking manifestation at the
macroscopic level under the form of the fractional Laplacian of Appendix 1.

To stress the main conclusion of the paper, we set aside the
non-Hamiltonian cases, and especially those where stochastic processes are
assumed to be already at work, whatever their origin might be. This is
where the microscopic dynamics are genuinely not differentiable, and
whether or not this lack of microscopic differentiability has macroscopic
manifestations depends on whether or not this non-differentiability can be
transmitted to the macroscopic level as a result of memory.  The present
paper sheds light also on this interesting, but less fundamental issue. A
much more important condition is that where the dynamics are described by a
Hamiltonian. In this case the microscopic dynamics are differentiable, and
in principle, there would be no microscopic randomness to transmit to the
macroscopic level.
        
However, one of the main tenets of the current literature on the
derivation of statistical mechanics from dynamics rests on the conviction
that randomness can be generated from hamiltonian dynamics either as a
consequence of the action of infinitely many degrees of freedom or as a
consequence of chaos. The Hamiltonian of Eq.(\ref{hami})
is an example of
the first kind, and the sense of the Van Hove limit in that case is that
the very short and the very large time-scales are beyond the range of
observation. Thus, randomness is a consequence of the observer's limitations.
The physical condition illustrated by Fig. 1, on the contrary, might be
related in some way to the case of chaos, in a sense that it
is convenient to properly discuss here.
        
Let us consider the well known case of a map resulting in
intermittency \cite{zum}, for instance, that introduced by Geisel and Thomae
\cite{geiselmap}. The motion under study is characterized by a long permanence in conditions of regular motion with bursts of chaotic dynamics concerning the
transition from one to the other laminar region. This is not a genuinely
Hamiltonian dynamics and the generator of intermittency is in fact a map.
According to the general discussion of \cite{zum}, similar properties can also
be exhibited by genuinely Hamiltonian systems such as the attractive model
of the egg-carton two-dimensional potential investigated by Geisel, Zacherl
and Radons \cite{geisel}. Another interesting model of the same type is the
three-dimensional Hamiltonian flow more recently studied by Zaslavsky,
Steven and Weitzner \cite{zsw93}. This paper, as well as that of Ref. \cite{geisel},
can be regarded as an example of Hamiltonian derivation of the L\'{e}vy
processes and consequently, according to the point of view adopted in this
paper, an example of a dynamical process described at the macroscopic level by
a fractional derivative in spite of its Hamiltonian, and consequently,
differentiable nature.

It has to be stressed, however, that the theory behind this macroscopic derivation should be applied to the numerical treatment of these processes: This theory does not have anything to do with the ideally exact solution of a dynamical model resting on a continuous treatment and the assumption of differentiability at any order. The numerical treatment is characterized by round-off errors and,
more importantly, by a discrete time representation, which forces the
system to depart from the conditions ideally established by its Hamiltonian
property. The intermittency of these systems are sufficient to create at
least temporary conditions of non-differentiability which are then
transmitted to the macroscopic level and changed, in the way described in
this paper, into fractional derivatives. The generalized version of the Van
Hove method serves the basic purpose of introducing conditions of
microscopic non-differentiability without leaving the theoretical treatment
and without entering the level of the numerical solution. In the case of
the dynamical approach to L\'{e}vy processes, the meaning of the generalized Van Hove method is closely related to the L\'{e}vy-Gnedenko generalized central
limit theorem \cite{gk54}. This is so because forcing the fluctuating variable
to become rigorously dichotomous imposes on the system enough coarse
graining as to produce randomness. We note also that in the case $1 < \beta < 2$, $T$ is proportional to the mean time duration the map spends in one laminar region.
Consequently if $t \gg T$, the number $N \equiv t/T$, which is very large, corresponds to the number of uncorrelated space transitions of intensity
$|x-x^{\prime}|$, with probability $1/ |x-x^{\prime}|^{1+\beta}$, made by the system. The L\'{e}vy-Gnedenko theorem insures that for $N$ tending to infinity a L\'{e}vy diffusion process is generated. From a physical point of view this has the same effect as forcing $T$ to zero.

We think that at this stage we are also in the right position to
establish an appealing connection with exciting results of the research
work of Refs. \cite{GASPARD98,SPOHN98}. Although some doubts are expressed by
the authors of Ref. \cite{SPOHN98} on the role of chaos to generate Brownian
motion (a role which might be sufficient but not necessary, as, in a sense, 
it is also shown by our chain model of Section II) we are inclined to believe
that the connection established by the authors of Ref. \cite{GASPARD98} 
between
Brownian motion and chaos is very attractive. However, it applies to the
condition of time-scale separation which is expected to generate ordinary
statistical mechanics. This paper is devoted, on the contrary, to studying
dynamical cases where this time-scale separation is missing, and
consequently, non ordinary statistical mechanics is generated. This might
generate the wrong impression that the method of analysis adopted in Ref.
\cite{GASPARD98}, based on the use of the Kolmogorov-Sinai entropy \cite{KS}, 
cannot
be applyed to the dynamical systems with no time-scale separation.
Actually, it has been recently shown \cite{TPZ97} that the Kolmogorov-Sinai
entropy can be generalized so as to be made efficient also in the case of
fractal dynamics. Furthermore, it was also recently shown \cite{condmat} that
the same entropic arguments naturally lead to the same form of spatial
fractional derivative as that of Eq. (\ref{frat}). This means that the 
Van Hove
generalized method discussed in this paper is expected to establish a
natural bridge between the non-extensive entropy of Tsallis 
\cite{CONSTANTINO88}
and the fractional derivative of Eq. (\ref{frat}) much in the same way as the
ordinary Van Hove method makes Hamiltonian dynamics compatible with the
standard diffusion equation of Eq. (\ref{mmm}), and so with ordinary extensive
thermodynamics behind it \cite{TSALLIS95}.

\section*{Acknowledgemnts}
One of the authors (A.R.) thanks INFM for partial support of this research.

\section*{Appendix 1}

We want to define the integrals and derivatives used in the fractional
calculus introduced in the text. First of all, let us recall the Riemann-Liouville definition of fractional
integral. Let us assume $\beta \ge 0$ and, following \cite{ross}, let us  define
the $\beta$-fractional integral of the function $f(x)$ to be 
\be
\frac{d^{-\beta }}{dx^{-\beta }}f(x)=\frac{1}{\Gamma (\beta )}\int_{c}^{x}\frac{f(y)dy}{(x-y)^{1-\beta }}.
\ee
We also define the $\beta $ fractional derivative of the function $f(x)$ as 
\be
\frac{d^{\beta }}{dx^{\beta }}f(x)=\frac{1}{\Gamma (n-\beta )}\frac{d^{n}}{%
dx^{n}}\int_{c}^{x}\frac{f(y)dy}{(x-y)^{\beta -n+1}},
\ee
where $n$ is the smallest integer larger than $\beta $, that is $n=[\beta ]+1$. The constant $c$ in the limits of the integrals is usually set to 0
(Riemann definition) or to $-\infty $ (Liouville definition). It is easy to
show that that for $\beta $ integer both definitions reduce to the ordinary
definitions of derivative and integral.

An equivalent definition makes use of Fourier Transform \cite{zava}. Let us
consider a function $f(x)$ with Fourier Transform $\hat{f}(k)$: 
\be
&&\hat{f}(k) = \frac{1}{\sqrt{2 \pi}} \int_{- \infty}^{\infty} f(x) e^{i k
x} dx, \\
&&f(x) = \frac{1}{\sqrt{2 \pi}} \int_{- \infty}^{\infty} \hat{f}(x) e^{- i k
x} dk.
\ee
The $n$-th derivative of $f(x)$ can be written as 
\be
D^{(n)} f(x) = {\cal F}^{-1} [(-ik)^{n} \hat{f}(k);x]  \label{base}
\ee
and a possible way of generalizing this expression to the $\beta$ derivative
of $f(x)$ is 
\be
D^{(\beta)} f(x) = {\cal F}^{-1} [(-ik)^{\beta} \hat{f}(k);x].
\label{antitr}
\ee
Eq.(\ref{antitr}) is equivalent to the convolution product: 
\be
D^{(\beta)} f(x) = \frac{1}{\sqrt{2 \pi}} \int_{- \infty}^{\infty}
d^{(\beta)}(x - y) f(y) dy,
\ee
where 
\be
d^{(\beta)}(x) \equiv \frac{1}{\sqrt{2 \pi}} \int_{- \infty}^{\infty}
(-ik)^{\beta} e^{- i k x} dk.  \label{z}
\ee
It is possible to calculate an explicit representation of the integral (\ref{z}) and the result is \cite{zava}: 
\be
&&D^{(\beta)}_+ f(x) = \frac{1}{\Gamma(n-\beta)} \frac{d^n}{dx^n}\int_{-\infty}^x \frac{f(y) dy}{(x - y)^{\beta-n+1}}  \label{dpiu} \\
&&D^{(\beta)}_- f(x) = - \frac{1}{\Gamma(n-\beta)} \frac{d^n}{dx^n}\int_x^{\infty} \frac{f(y) dy}{(x - y)^{\beta-n+1}},
\ee
where again $n = [\beta] + 1$ and the $+(-)$ corresponds to evaluating the
integral (\ref{z}) in the upper (lower) complex $k$ plane. It is apparent
that Eq.(\ref{dpiu}) coincides with the Liouville definition of fractional
derivative: 
\be
D^{(\beta)}_+ f(x) \equiv \frac{d^{\beta}}{d x^{\beta}} f(x).
\ee
A different way of generalizing Eq.(\ref{base}) is the following. Let us start
from the definition of second derivative 
\be
D^{(2)} f(x) = {\cal F}^{-1} [(-k^2) \hat{f}(k); x]
\ee
and let us generalize it as: 
\be
{\cal D}^{(2 \beta)} f(x) = {\cal F}^{-1} [(-k^2)^{\beta} \hat{f}(k);x].  \label{antitra}
\ee
Eq.(\ref{antitra}) is equivalent to the convolution product: 
\be
{\cal D}^{(2 \beta)} f(x) = \frac{1}{\sqrt{2 \pi}} \int_{-\infty}^{\infty} c^{(\beta)}(x - y) f(y) dy  \label{convi}
\ee
where 
\be
c^{(\beta)}(x) \equiv \frac{(-1)^{\beta}}{\sqrt{2 \pi}} \int_{-\infty}^{\infty} |k|^{2 \beta} e^{- i k x} dk.  \label{zz}
\ee
If 
\be
0 < \beta < 1,
\ee
the integral (\ref{zz}) can be evaluated explicitly \cite{grad} 
\be
c^{(\beta)}(x) = \frac{(-1)^{\beta}}{\sqrt{2/ \pi} \; \Gamma(-2 \beta) \cos
(\beta \pi)} \frac{1}{|x|^{2 \beta + 1}},
\ee
and setting $2 \beta = \mu$ ($\Rightarrow 0 < \mu < 2$), we get: 
\be
\frac{1}{|x|^{\mu + 1}} = - \frac{2 b (-1)^{\mu /2}}{\mu (\mu - 1) Q}
c^{(\mu)}(x).  \label{unosu}
\ee
Using now Eq.(\ref{condi}), Eq.(\ref{convi}) and Eq.(\ref{unosu}), {Eq.(\ref{sw}) can
be expressed in terms of the fractional derivative ${\cal D}^{(\mu)}$.
This gives Eq.(\ref{frat}). Notice that Eq.(\ref{frat}) is consistent with Eq.(\ref{swf}) since 
\be
{\cal F} [c^{(\mu)}(x);k] = (-k^2)^{\mu /2} = i^{\mu} |k|^{\mu}.
\ee
}

\section*{Appendix 2}

Our aim is now to show that the conventional Van Hove limit produces in the
case of exponential relaxation for the correlation function the same result
as the traditional approach. Let us consider the following equation of motion for the probability density $p(x,t)$: 
\be
\frac{\partial}{\partial t} p(x,t) = \langle v^2 \rangle_{eq} \int_0^t
\Phi_{v}(t-t^{\prime}) \frac{\partial^2}{\partial x^2} p(x,t^{\prime})
dt^{\prime}.  \label{eqv}
\ee
Eq.(\ref{eqv}) is the same as Eq.(\ref{nongdiff}) and the bath variable, now
denoted with $v$, is not dichotomous. Assuming 
\be
\Phi_{v}(t) = e^{- \gamma t}, \;\;\;\;\;\; \gamma = \frac{1}{\tau} \label{assun}
\ee
and differentiating both sides of Eq.(\ref{eqv}), we obtain 
\be
\frac{\partial^2}{\partial t^2} p(x,t) = - \gamma \frac{\partial}{\partial t}
p(x,t) + \langle v^2 \rangle_{eq} \frac{\partial^2}{\partial x^2} p(x,t), \label{ufra}
\ee
whose Fourier Transform is given by 
\be
\frac{\partial^2}{\partial t^2} \hat{p}(k,t) + \gamma \frac{\partial}{\partial t} \hat{p}(k,t) + \langle v^2 \rangle_{eq} k^2 \hat{p}(k,t) = 0.
\label{eqvf}
\ee
Both Eq.(\ref{ufra}) and (\ref{eqvf}) correspond to the telegraphic equation. Considering the condition 
\be
\dot{\hat{p}}(k,0) = 0,
\ee
the exact solution of (\ref{eqvf}) is written as 
\be
\hat{p}(k,t) = A \left[e^{\alpha_+(k)t} - \frac{\alpha_+(k)}{\alpha_-(k)}
e^{\alpha_-(k)t}\right]  \label{ex}
\ee
with $A$ to be specified according to the normalization condition and $\alpha_{\pm}$ given by 
\be
\alpha_{\pm} = - \frac{\gamma}{2} \pm \frac{\gamma}{2} \sqrt{1 - \frac{4
\langle v^2 \rangle_{eq} k^2}{\gamma^2}}.
\ee
Assuming now 
\be
\frac{4 \langle v^2 \rangle_{eq} k^2}{\gamma^2} \ll 1  \label{condir}
\ee
the solution (\ref{ex}) can be expanded in a Taylor series, giving 
\be
\hat{p}(k,t) & = & A \left[\exp\left(-\frac{\langle v^2 \rangle_{eq} k^2}{\gamma}t \right) \right. \nonumber \\
&& \qquad \left. - \frac{\langle v^2 \rangle_{eq} k^2/ \gamma}{\gamma - \langle v^2 \rangle_{eq} k^2/ \gamma} \exp \left(- \gamma t + \frac{\langle v^2 \rangle_{eq} k^2}{\gamma} t \right) + {\cal O}\left(\frac{\langle v^2 \rangle_{eq}^2}{\gamma^3} \right) \right].  \label{ordini}
\ee
Note that setting condition (\ref{condir}) is compatible with the Van Hove limit resting on setting $\gamma \rightarrow \infty$. At this stage we have to apply to Eq. (\ref{ordini}) the Van Hove limit, which in addition to setting $\gamma \rightarrow \infty$ rests also on making the limit $\langle v^2 \rangle_{eq} \rightarrow \infty$ in such a way that
\be
D = {\rm Lim_{VH}} \frac{\langle v^2 \rangle_{eq}}{\gamma}.
\ee
We note that this procedure makes the terms ${\cal O}\left(\frac{\langle v^2 \rangle_{eq}^2}{\gamma^3} \right)$ disappear so as to recover the ordinary diffusion equation solution: 
\be
\hat{p}(k,t) = A e^{-D k^2 t}.
\ee
On the other hand, the same result can be obtained by applying the Van
Hove limit directly to the equation of motion (\ref{eqv}) with the
assumption (\ref{assun}) and also making use of both Eq.(\ref{tr1}) and Eq.(\ref{tr2}).
This proves that ordinary Brownian diffusion rests on the Van Hove limit.


\begin{thebibliography}{99}
\bibitem{nota}  In our opinion the meaning of the Van Hove method has been
made transparent by the discussion of Zwanzig in Ref. \cite{zwa}. For this
reason, we refer ourselves to the work of Zwanzig more than to the original
paper by Van Hove (Ref. \cite{vh}).
\bibitem{note} It is important to stress that the fractional time derivative
adopted in \cite{MKS98} aims at a purpose different from that of this section.
On the contrary, the fractional time derivative adopted in \cite{MN98},
equivalent to setting $T = 0$ in the denominator of $T^{\beta}/(T^2+t^{\prime 2})^{\beta/2}$ in Eq. (\ref{intprop}), conflicts with the generalized Van Hove method and
consequently with the numerical and theoretical conclusions of both Refs.
\cite{ZK93} and \cite{alle}.
\end{thebibliography}
\end{document}